\documentstyle{article}
\font\tenbf=cmbx10
\font\tenrm=cmr10
\font\tenit=cmti10
\font\elevenbf=cmbx10 scaled\magstep 1
\font\elevenrm=cmr10 scaled\magstep 1
\font\elevenit=cmti10 scaled\magstep 1

\renewenvironment{thebibliography}[1]
 { \elevenrm
   \begin{list}{\arabic{enumi}.}
    {\usecounter{enumi} \setlength{\parsep}{0pt}
     \setlength{\itemsep}{3pt} \settowidth{\labelwidth}{#1.}
     \sloppy
    }}{\end{list}}

\begin{document}
\begin{center}{{\tenbf MAXWELL-CHERN-SIMONS CASIMIR EFFECT\\}
\vglue 1.0cm
{\tenrm KIMBALL A. MILTON \\}
\baselineskip=13pt
{\tenit  Department of Physics and Astronomy, University of
Oklahoma\\}
\baselineskip=12pt
{\tenit Norman, OK 73019-0225, USA\\}
\vglue 0.8cm
{\tenrm ABSTRACT}}
\end{center}
\vglue 0.3cm
{\rightskip=3pc
 \leftskip=3pc
 \tenrm\baselineskip=12pt
 \noindent
In odd-dimensional spaces, gauge invariance permits a Chern-Simons
mass term
for the gauge fields in addition to the usual Maxwell-Yang-Mills
kinetic
energy term.  We study the Casimir effect in such a (2+1)-dimensional
Abelian
theory.  For the case of parallel conducting lines the result is the
same
as for a scalar field.  For the case of circular boundary conditions
the results
are completely different, with even the sign of the effect being
opposite
for Maxwell-Chern-Simons fields and scalar fields. We further examine
the
effect of finite temperature. The Casimir stress is found to be
attractive
at both low and high temperature.
  Possibilities of observing this effect in the
laboratory are discussed.

\vglue 0.6cm}
{\elevenbf\noindent 1. Introduction
}
\vglue 0.2cm
\baselineskip=14pt
\elevenrm

By now it is well-known that, for theories in odd-dimensional
spaces, one can add a gauge invariant   Chern-Simons mass term
for the gauge field in addition to the usual Maxwell-Yang-Mills
term. Recently there has
been considerable interest in such a $(2+1)$-dimensional
Abelian theory in connection with the studies of the fractional
quantum Hall effect
in semiconductors and of high-$T_c$ superconductivity
in copper-oxide crystals.  The Lagrangian for the
Maxwell-Chern-Simons theory written in
curvilinear coordinates is
\begin{equation}
{\cal L}=-\sqrt{-g}{1\over4}F^{\mu\nu}F_{\mu\nu}
+{1\over4}\mu\epsilon^{\mu
\alpha\beta}F_{\alpha\beta}A_\mu,\label{eq:lag}
\end{equation}
where $g$ is the determinant of the metric $g_{\mu\nu}$ and
$\epsilon^{\mu\alpha\beta}=\sqrt{-g}e^{\mu\alpha\beta}$
is a tensor density, with $\epsilon^{012}=1$.
In terms of the dual tensor
$F^\lambda={1\over2}e^{\lambda\alpha\beta}F_{\alpha\beta}$
we can rewrite (\ref{eq:lag}) as
\begin{equation}
{\cal L}={1\over2}\sqrt{-g}(F^\lambda F_\lambda
+\mu F^\lambda A_\lambda)
.\label{eq:lagp}
\end{equation}
Varying ${\cal L}$ with respect to $A_\mu$ we find the
equations of motion
\begin{equation}
\epsilon^{\mu\alpha\beta}\partial_\alpha F_\beta+\mu
\sqrt{-g}F^\mu=0.
\label{eq:eom}
\end{equation}

We solve these equations subject to perfect conductor boundary
conditions.  That is, the tangential electric
field must vanish on the boundary, or the normal component of
the dual field vanishes, $F_n=0$.
 It is interesting to note that this is precisely the
condition necessary to ensure the gauge invariance of
the Lagrangian (\ref{eq:lagp}).
That is, the mass term in (\ref{eq:lagp})
is gauge invariant only if we neglect the surface term
\begin{equation}
{1\over2}\mu\int dx\, \sqrt{-g}
F^\lambda\partial_\lambda\Lambda=
{1\over2}\mu\int dS_\lambda \sqrt{-g} F^\lambda
\Lambda=0,
\end{equation}which is true if the normal component of $F^\lambda$
vanishes on the bounding surfaces.

\vglue 0.6cm
{\elevenbf\noindent 2. Green's Functions}
\vglue 0.4cm

The stress tensor density following from (\ref{eq:lagp}) is
\begin{equation}
t^{\alpha\beta}=\sqrt{-g}(F^\alpha F^\beta
-{1\over2}g^{\alpha\beta}
F_\lambda F^\lambda).\label{eq:stresstensor}
\end{equation}
We introduce the Green's function
$G_{\mu\nu}$ according to
\begin{equation}
F_\mu(x)=\int dx'\,\sqrt{-g(x')}G_\mu{}^\nu(x,x')
J_\nu(x').\label{eq:green}
\end{equation}
The equations of motion (\ref{eq:eom}) imply that
$G_{\mu}{}^{\nu}$ satisfies the
equation
\begin{equation}
e_\mu{}^{\nu\lambda}\partial_\nu G_\lambda{}^\alpha
+\mu G_\mu{}^\alpha
=-{1\over\sqrt{-g}}g_\mu{}^\alpha\delta(x-x'),
\label{eq:geq}
\end{equation}
where $e_\mu{}^{\nu\lambda}=g_{\mu\beta}
 \epsilon^{\beta\nu\lambda}/\sqrt{-g}$.
In terms of vacuum expectation values of fields
\begin{equation}
G_{\mu\nu}(x,x')=i\langle F_\mu(x) A_\nu(x')\rangle,
\label{eq:gfa}
\end{equation}
and so
\begin{equation}
\langle t^{\alpha\beta}\rangle=\lim_{x'\to x}
{1\over i}(\epsilon^{\beta\gamma
\sigma}\partial'_\gamma G^\alpha{}_\sigma
-{1\over2}g^{\alpha\beta}g_{\mu\nu}
\epsilon^{\nu\gamma\sigma}\partial'_\gamma
G^\mu{}_\sigma),
\end{equation}
where the limit $x'\to x$ is to be taken symmetrically.

\vglue 0.5cm
{\elevenbf \noindent 3. Perfectly Conducting Parallel Lines}
\vglue 0.4cm

For the case of parallel conducting lines, we can introduce a
transverse spatial Fourier transform together with a Fourier
transform in time:

\begin{equation}
G^{\mu\nu}(x,x')=\int{d\omega\over2\pi}e^{-i\omega(t-t')}
\int{dk\over2\pi}e^{ik(y-y')}{\cal G}^{\mu\nu}(k,\omega;x,x').
\end{equation}
It is straightforward to solve for the components ${\cal
G}^{\mu\nu}$.
The details are given in Ref.~\cite{mcsce1}.  The result for the
stress
tensor on the (inside of) the bounding surfaces is
\begin{equation}
t^{11}(0\mbox{ or } a)
={i\kappa\over2}\cot\kappa a,
\end{equation}
where $\kappa^2=\omega^2-k^2-\mu^2$.  To find the force per unit
length,
we compute the discontinuity of $t^{11}$ across the boundary,
and integrate over transverse momentum and frequency, after
performing
a Euclidean rotation:
\begin{equation}
f=-{1\over16\pi a^3}\int_{2\mu a}^\infty dy
{y^2\over e^y-1}\to-{\zeta(3)\over8\pi a^3}, \quad\mbox{as }\mu
a\to0.\label{eq:para}
\end{equation}

This result, and the corresponding non-zero temperature result,  are
exactly the same as for a scalar field in the
same geometry.

\vglue 0.5cm
{\elevenbf \noindent 4. Circular Boundary Conditions \hfil}
\vglue 0.4cm

In this case we introduce the Fourier transform appropriate to
the polar coordinates:
\begin{equation}
G_\mu{}^\nu(x,x')=\int{d\omega\over 2\pi}
e^{-i\omega(t-t')}\sum_{m=-\infty}
^\infty e^{im(\theta-\theta')}{\cal G}_\mu{}^\nu (r,r').
\end{equation}
Here,  the calculation is considerably more involved,
so we will content ourselves with presenting numerical results
at $\mu=0$ and $T=0$:  The total Casimir force on the circle can
be broken into two pieces, coming from $m=0$ and $m\ne0$ terms:
$F=F_0+F_{\rm LT}$, where
\begin{equation}
F_0=-{1\over2\pi a^2}\int_0^\infty dx\, x{d\over dx}
\ln [2x I_1(x)K_1(x)]=-{0.254\over a^2}\label{eq:mu0m0}
\end{equation}
and
\begin{equation}
F_{\rm LT}=-{1\over4\pi^2 a^2}\int_0^\infty
{dy \,y\over e^y-1}=-{1\over24a^2}.
\label{eq:lt}
\end{equation}
The corresponding scalar result is six times smaller and opposite in
sign.
The force falls off rather rapidly with $\mu a$, being only $1/2$
the maximum value at $\mu a = 2$.  For a discussion of this and
the temperature dependence, the reader is referred to
Ref.~\cite{mcsce2}.

\vglue 0.5cm
{\elevenbf \noindent 5. Discussion \hfil}
\vglue 0.4cm
To get an idea of the magnitude of this effect, consider a circle of
radius
100 \AA.  According to (\ref{eq:mu0m0}), the corresponding Casimir
energy is
only about 5 eV.  One could imagine a device consisting of a great
many
circular regions, so the coherent effect could be much larger.  A
much
larger effect occurs for parallel lines as well: for two lines
separated
by 100 \AA, the energy/length is about $4\times 10^5$ eV/cm.  One
must
further bear in mind that the Chern-Simons field describing anyons is
topological, and should not be confused with a observable Maxwell
field;
our speculation is that in two-dimensional confined geometries, the
photon may acquire a topological mass, since such is not prohibited
by
gauge invariance.  It remains a challenge to the experimenter to
find a signature of this effect.  We hope we have sown seeds of
inspiration.
\vglue 0.5cm
{\elevenbf \noindent 5. Acknowledgements \hfil}
\vglue 0.4cm
We thank the US Department of Energy for partial financial support.
\vglue 0.5cm
{\elevenbf\noindent 6. References \hfil}
\vglue 0.4cm

\vglue 0.5cm
\end{document}